\documentclass[runningheads,a4paper]{llncs}
\usepackage{amssymb} 
\usepackage{graphicx}
\usepackage{url}
\usepackage{float}
\usepackage{soul,color}  
\usepackage{amsmath}

\begin{document}

\title{Joining Jolie to Docker}

\subtitle{Orchestration of Microservices on a Containers-as-a-Service Layer}

\author{
Alberto Giaretta\inst{1},
Nicola Dragoni\inst{1,2} and
Manuel Mazzara\inst{3}
\institute{Centre for Applied Autonomous Sensor Systems, \"{O}rebro University, Sweden
\and DTU Compute, Technical University of Denmark, Denmark
\and Innopolis University, Russian Federation}}

\toctitle{Lecture Notes in Computer Science}
\tocauthor{Authors' Instructions}
\maketitle

\begin{abstract}
Cloud computing is steadily growing and, as IaaS vendors have started to offer pay-as-you-go billing policies, it is fundamental to achieve as much elasticity as possible, avoiding over-provisioning that would imply higher costs. In this paper, we briefly analyse the orchestration characteristics of PaaSSOA, a proposed architecture already implemented for Jolie microservices, and Kubernetes, one of the various orchestration plugins for Docker; then, we outline similarities and differences of the two approaches, with respect to their own domain of application. Furthermore, we investigate some ideas to achieve a federation of the two technologies, proposing an architectural composition of Jolie microservices on Docker Container-as-a-Service layer.
\end{abstract}

\section{Introduction}
\label{sec:intro}
As the cloud computing paradigm keeps gaining consensus nowadays, a smart and easy way to provide distributed services is of utmost importance. Furthermore, the new pay-as-you-go billing policies~\cite{IHJ}, offered by vendors such as Amazon EC2~\cite{EC2}, boost the requirement of efficient service orchestration tools, since inefficient management of resources entails a higher economic burden for business companies.

Before the cloud revolution, Virtual Machines have been the standard envelope for distributed services, but their conservative approach towards resource management, along with their intrinsic provisioning of a whole-functioning machine, makes them too much wasteful with respect to their actual necessities. As an example, deploying a simple web-server instance within a VM implies that a complete machine is given, with all its own layers, which means over-provisioning by design.

Therefore, a new composition approach is needed in order to achieve a federation of infrastructures, along with as much elasticity as possible.

\section{Service Orchestration}
\label{sec:service_orch}
Before cloud computing, software applications have traditionally been monolithic \cite{Dragoni2017}. Thus, developing a monolithic software implied, by design, that communications between components were always possible, being all the parts hosted on the same machine.

In a cloud world some of previous certainties, such as the components reachability, do not hold. Components of a complex software could be scattered around the cloud, meaning that communications problems could arise, like high delays, high jitter or even total lack of network connection~\cite{WN,Weinman}. Furthermore, load requirements are not static and resources need to be managed dynamically, accordingly to the real necessities: this is where the concept of \textit{service orchestration} arises. Service orchestration~\cite{KACWE} could be interpreted as the automatic provision and release of resources, whether virtual or physical, necessary to deliver the agreed service level. 

While old monolithic software required \textit{vertical scaling} to alleviate resources bottlenecks (i.e., improvement of the current machine hardware), \textit{scaling out}, even known as \textit{horizontal scaling}, is the most important characteristic of cloud computing~\cite{KACWE,Dragoni2017b}. Instead of scaling vertically (which can be really expensive, if higher-end hardware is needed), with horizontal scaling additional machines are used, and the underperforming services are replicated in order to improve the overall services' performance. Furthermore, if the currently available resources are enough but unbalanced, graceful ways to pause, migrate and restart services must be given, to achieve the capability to rearrange them and optimize the resources. Last, but not least, it is essential to stop the additional services once they are no longer needed, otherwise pay-as-you-go billings would become uselessly encumbering.

Therefore, it is easy to see that complex problems come to surface in a cloud computing architecture. Services need to be movable, among the other things, to achieve elasticity, and this movability leads to other non-trivial problems. A service orchestrator, being the component that handles the running services to ensure that stipulated SLAs are met, should~\cite{KACWE}:
\begin{itemize}
\item Replicate services;
\item Migrate services;
\item Start services;
\item Pause services;
\item Terminate services.
\end{itemize}

\section{Jolie}
\label{sec:jolie}
Two main approaches exist to write distributed software: creating a library (or a framework) that adds up to an already existing language, or creating a new service-oriented programming language. \textit{Jolie}~\cite{JOLIE}, acronym for \textit{Java Orchestration Language Interpreter Engine}, is a completely new microservice programming language with a large supporting community, both academic and industrial \cite{Bandura16}. Based upon a \textit{C}-like syntax, it is the attempt to simplify the software development by overcoming the complexity of other existing languages like \textit{BPEL}, which are hardly comprehensible to humans due to their \textit{XML}-like syntax~\cite{MGLZ}. Specifically created to write microservices, it supports this idea at the level of the foundational primitives \cite{Guidi2017}. One of the peculiar strengths is the separation between behaviour (what the service does) and deployment (how the service connects with the outside world). 

Jolie is the only language that natively supports the microservice paradigm \cite{Guidi2017}. Although workflow engines are not a novelty \cite{maurer1999software}, and languages to describe service orchestration existed before \cite{WS-BPEL}, Jolie has been designed with fine-grained procedural constructs in order not only to provide high-level orchestration, but to program the internal logic of a single microservice.

While microservices are inherently suitable to develop cloud-oriented software, Jolie in its current version lacks of service orchestration features (e.g., the capability of scaling out and migrating services), which means that it is far from being appropriate for real-life cloud applications. Basic features have been implemented, such as service discovery, but it is not enough. Ideally, a software developer should be able to write and deploy microservices having no clue about the network framework because components displacement it is likely to change many times: as an example, a developer should not have to specify the IP address of the service discovery server.

To obtain service orchestration, a SOA-based architecture called PaaSSOA has been proposed and implemented for Jolie~\cite{GAV,BZV}. Among the various characteristics of PaaSSOA, the most important one for our work is SOABoot, which is a sort of container for Jolie services. The SOABoots altogether form the Service Container layer, exposed at SaaS and PaaS level.

A SOABoot can receive services implementations, store, activate and deactivate them. This clearly means that elasticity is obtainable, because the PaaSSOA Scheduler is able to request new VMs to the IaaS level~\cite{GAV} (every one with its own SOABoot instance), migrate Jolie services between different VMs and even start/stop them. The strong point of the PaaSSOA approach is that, except when new VMs are needed, all the arrangements are strictly done at PaaS level.

Every PaaSSOA VM automatically provisions a SOABoot instance. Therefore, elasticity is obtained by design, simply increasing or decreasing the number of running VMs, within which Jolie services are able to execute.

To obtain all these things, PaaSSOA provides a set of \textit{functions} called \textit{Service Deployer and Monitor (SDM)} which delivers: \textit{deployment}, to migrate or deploy the services; \textit{scheduler}, to schedule the needed deployment, accordingly to the available resources; \textit{negotiator}, to negotiate resources with the IaaS, compatibly with the \textit{Service Level Agreement (SLA)} stipulated beforehand; \textit{monitor}, to check the SLA conformance and take actions in case of unmet SLA. Generally speaking, every PaaS layer should provide these characteristics.

With regard to the desirable characteristics of a service orchestrator, described in Section \ref{sec:service_orch}, it looks crystal clear that Jolie alone is unable to supply service orchestration in a cloud environment, which is a huge shortcoming for a service-oriented language that aspires to be suitable for the cloud. Nonetheless, Jolie paired with PaaSSOA fully satisfies the expressed requirements in Section \ref{sec:service_orch}.

\section{Docker}
\label{sec:docker}
Docker~\cite{Docker} is an open-source software that deploys software applications within software containers. Even though, at first sight, this has been done for many years with virtual machines, VMs aim to deliver to the final user a simulation of a complete machine, and this completeness comes with a price, in form of heaviness and required resources~\cite{HGGWGH}. The intuition behind the containers concept is to package only the strictly necessary parts (e.g., not the OS kernel) and enable the guest to use the underlying layers, lent by the host, instead of simulating them. Investigations have shown that containers can match, and even outdo, VMs from a performance point of view~\cite{FFRR}.

Docker actually achieves container orchestration by using orchestrator plugins, such as Kubernetes, which can effortlessly and transparently start, stop and move containers around the cloud~\cite{B}. Kubernetes, for instance, can monitor and manage containers in many ways. It is able to launch new containers in already-existing VMs, to migrate containers from a VM to another one and even communicate with the IaaS, in order to obtain the provisioning of new VMs, within containers can boot. Furthermore, Kubernetes gives the opportunity to create \textit{pods}, which are logical sets of containers, and everything can whether be hosted within VMs or bare metal machines.

The strength of Docker is the implicit promise of delivering PaaS functionalities with an extremely simplified mechanism, becoming a standard that avoids vendor lock-ins and permits easy multi-providers cloud solutions. If Docker imposes itself, developers could easily load their software on the containers, wherever they are hosted, eliminating all the struggle with APIs and tools, which are specific for each IaaS provider~\cite{DRK}. All of this is possible by introducing an additional layer, which is called \textit{Containers-as-a-Service (CaaS)}, into the cloud computing stack which fits between the IaaS and the PaaS and that, ideally, should be the same for all the IaaS providers.

As like as PaaSSOA, Docker equipped with Kubernetes (or a similar orchestrator plugin) totally satisfies the requirements exposed in Section \ref{sec:service_orch}, achieving a full-scale level of service orchestration.

\section{Comparison Between Jolie and Docker}
Section \ref{sec:jolie} and Section \ref{sec:docker} show, respectively, the main characteristics of Jolie and Docker with regards to service orchestration within the cloud. Interestingly, we can draw an analogy between the duo Jolie/PaaSSOA and Docker/Kubernetes, even though they exist and operate at different layers.

First of all, both Kubernetes and PaaSSOA are capable of communicating with the IaaS layer as needed, to ask for new resources. Furthermore, both of them can start, stop and move services within the cloud. We can even envision a strong similarity between Kubernetes pods and PaaSSOA Service Container, in their logical wrapping of services operating on different machines.

The main difference between the two approaches, is that PaaSSOA can move services from a SOABoot to another one without involving the IaaS layer~\cite{GAV}, if the available resources are enough, whereas Docker is tied to deal with the IaaS every time that a migration is needed.

\section{Federation of Jolie and Docker}
Even though the combination of Jolie and PaaSSOA is able to deliver service orchestration in a cloud world as Docker and Kubernetes do, it does not mean that one solution should exclude the other. As a matter of fact, they could cooperate to achieve a cogent service orchestration spread over their respective layers of application. Therefore, we expose three main ideas on how containers could be included into a PaaSSOA solution:
\begin{itemize}
\item With respect to the SOABoot original architecture~\cite{GAV}, simply substitute VMs with containers, managing the orchestration at the PaaS layer;
\item Fix each service into its own container and trust Docker to deal with orchestration tasks;
\item Substitute VMs with containers and tweak PaaSSOA to communicate with Docker.
\end{itemize}

\begin{figure}[t]
\center
\includegraphics[trim=10mm 10mm 10mm 10mm,clip,width=0.90\textwidth]{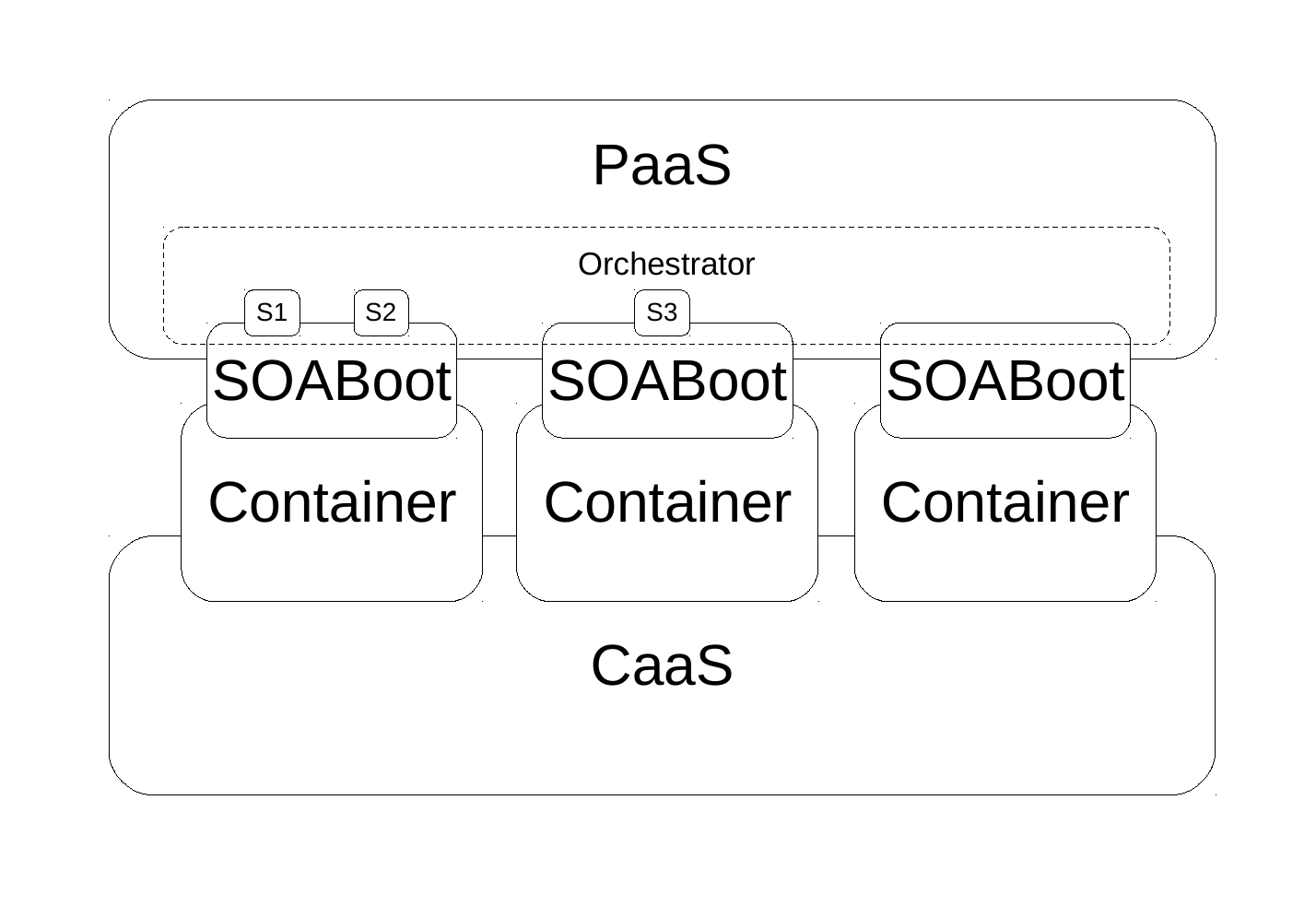}
\caption{The first solution proposed totally relies upon the PaaS layer, in order to achieve service orchestration. Services (e.g., S1) are the handling units.}
\label{fig:PaaS_Orch}
\end{figure}

The first approach, shown in Figure~\ref{fig:PaaS_Orch}, is very simple. Containers are lighter than VMs, therefore this would result in a decrease of overprovisioned resources. At the same time, this solution totally relies on service orchestration at the PaaS level, without manipulating containers at the CaaS level.

\begin{figure}[t]
\center
\includegraphics[trim=10mm 10mm 10mm 10mm,clip,width=0.90\textwidth]{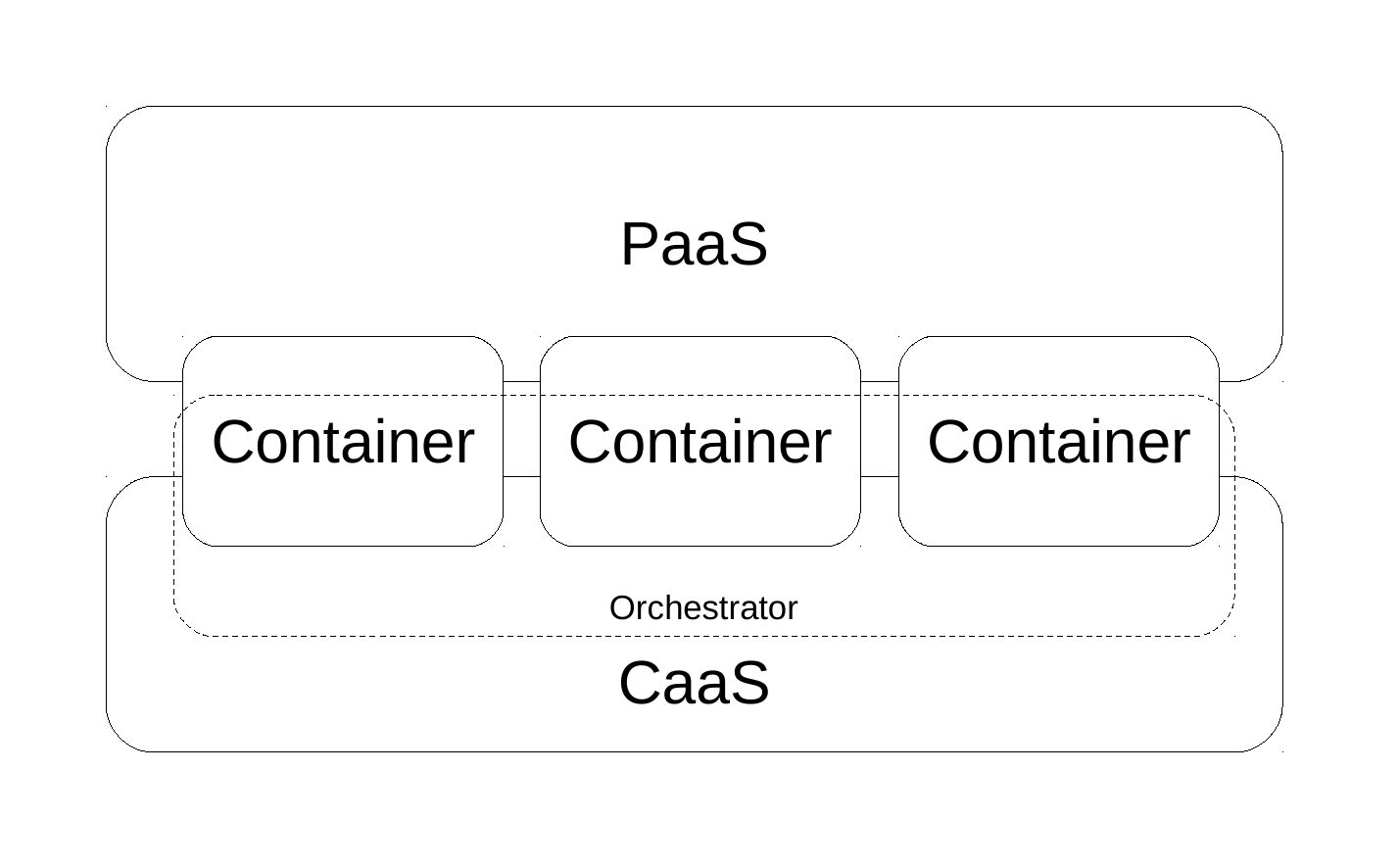}
\caption{The second solution proposed totally relies upon the CaaS layer, in order to achieve service orchestration. Containers are the handling units.}
\label{fig:CaaS_Orch}
\end{figure}

The second approach slims down the PaaS layer involvement. A predefined set of services is fixed in its own Docker container, therefore all the elasticity is achieved at CaaS level and SOABoot is no longer needed. On the one hand, this solution has the great virtue to manage every service, whether it is a Jolie service or not, in the same way at the CaaS level, being containers the handling units used. On the other hand, the PaaS layer is totally deprived of its own service orchestration tasks, and the whole architecture loses the capability to handle services at a finer detail (i.e., services at the PaaS level are no more manipulable). Figure~\ref{fig:CaaS_Orch} shows our proposal.

\begin{figure}[t]
\center
\includegraphics[trim=10mm 15mm 10mm 15mm,clip,width=0.90\textwidth]{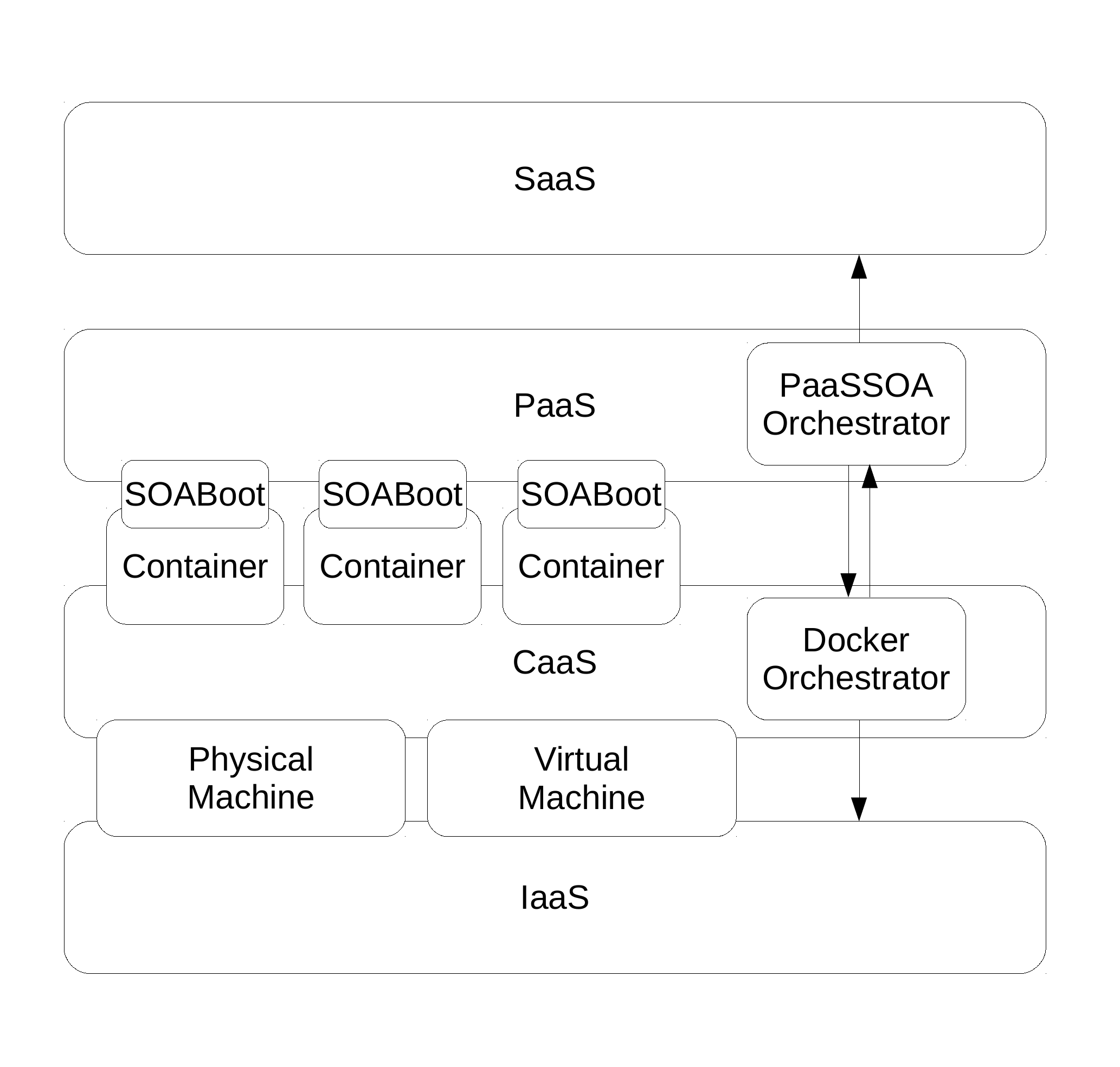}
\caption{The architectural point of view of PaaSSOA on the shoulders of Docker, where service orchestration is done at both PaaS and Caas layers. In this scenario, both services and containers are handling units.}
\label{fig:PaaS_CaaS}
\end{figure}

The third approach is the most complex and the most flexible of all the three. The idea is to keep all the characteristics of PaaSSOA and Docker, enabling PaaSSOA to communicate with Docker orchestrator, trying to find a trade-off between the requirements of SaaS and IaaS layers. The CaaS layer, introduced by Docker, fits between the IaaS and the PaaS and includes the Docker orchestrator, while the PaaS layer includes PaaSSOA. Our architectural view is shown in Figure~\ref{fig:PaaS_CaaS}.

Using this approach, the federation of orchestrators would have four different options to attain a balanced set of services:
\begin{itemize}
\item Ask Docker for a new container, if resources are scarce;
\item Rearrange services at PaaS level, without involving the underlying CaaS, if resources are enough but services are unbalanced;
\item Entrust Docker to reorganize containers at CaaS level, if resources are enough but services are unbalanced.
\end{itemize}

In particular, the capability of the PaaS and CaaS layers to dialogue seems fundamental to obtain an agreement between PaaSSOA and Docker orchestration requirements. Two load balancing components that do not communicate, quite certainly do not share the same point of view on balance, and this different point of view would lead to undesirable episodes of two components fighting each other, constantly trying to achieve their own concept of balance.

\section{Conclusion}
In this paper, we have analysed the concept of service orchestration in a cloud computing scenario. Then, we have inspected how service orchestration is done with Jolie, a microservices programming language, and Docker, an automatic deployer of applications within containers. Furthermore, we have drawn some analogies between the two different worlds and, most importantly, we have proposed an architectural solution to join the best of the two worlds to achieve an elastic and fine grained constellation of services.

Our research team has worked on the microservice paradigm since the early stages of its industrial adoption and cooperated with large companies in the process of migration \cite{DDLM2017}. Several projects have been conducted relying on the Jolie programming language \cite{Salikhov2016a,Salikhov2016b}, as well as covering the development of parts of the language itself (extension of the type system \cite{Safina2016}, prototyping of static type checking \cite{Tchitchigin16}, addition of more iterative control structures to support programming, and inline automatic documentation \cite{Bandura16}).
Often Jolie and Docker have been compared and we have often been asked why we chose one instead of the other. Therefore, future steps of the research, and of the adoption of the microservice paradigm, should focus  on the experimentation of the architectural solution proposed in this paper that promises to combine the best of Jolie and Docker.
Other software development projects with a strong emphasis on distribution and componentization could greatly benefit from a reorganization of the software architecture, for example distributed social networks \cite{MBGDMQN2013}.

%\bibliographystyle{IEEEtran}

\end{document}